\documentclass[journal]{IEEEtran}
\usepackage{array}
\usepackage{url}
\usepackage{amsmath}
\usepackage{amssymb}
\usepackage{graphicx}
\usepackage{cite}

\makeatletter

\newcommand{\noun}[1]{\textsc{#1}}
\providecommand{\tabularnewline}{\\}
\@ifundefined{showcaptionsetup}{}{%
 \PassOptionsToPackage{caption=false}{subfig}}
\usepackage{subfig}
\makeatother

\begin{document}

\title{The Accuracy-Privacy Tradeoff \\of Mobile Crowdsensing}

\author{Mohammad~Abu~Alsheikh, Yutao Jiao, Dusit~Niyato, \\Ping Wang, Derek Leong, and Zhu Han
\thanks{M.~Abu~Alsheikh, Y. Jiao, D.~Niyato,  and P.~Wang are with the School of Computer Science and Engineering, Nanyang Technological University, Singapore 639798 (mohammad027@e.ntu.edu.sg, yjiao001@e.ntu.edu.sg,  dniyato@ntu.edu.sg, and wangping@ntu.edu.sg). D.~Leong is with the Institute for Infocomm Research, Singapore 138632 (dleong@i2r.a-star.edu.sg). Z.~Han is with the School of Electrical and Computer Engineering, University of Houston, TX, USA 77004 (zhan2@uh.edu).}
\thanks{Contact: \textbf{D. Niyato}, School of Computer Engineering, Nanyang Technological University, Block N4-02a-32, Nanyang Avenue, Singapore 639798, Tel:~+65-6790-4121, Fax:~+65-6792-6559, Email: dniyato@ntu.edu.sg.}
}
\maketitle
\begin{abstract}
Mobile crowdsensing has emerged as an efficient sensing paradigm which combines the crowd intelligence and the sensing power of mobile devices, e.g.,~mobile phones and Internet of Things (IoT) gadgets. This article addresses the contradicting incentives of privacy preservation by crowdsensing users and accuracy maximization and collection of true data by service providers. We firstly define the individual contributions of crowdsensing users based on the accuracy in data analytics achieved by the service provider from buying their data. We then propose a truthful mechanism for achieving high service accuracy while protecting the privacy based on the user preferences. The users are incentivized to provide true data by being paid based on their individual contribution to the overall service accuracy. Moreover, we propose a coalition strategy which allows users to cooperate in providing their data under one identity, increasing their anonymity privacy protection, and sharing the resulting payoff. Finally, we outline important open research directions in mobile and people-centric crowdsensing.
\end{abstract}

\begin{IEEEkeywords}
Privacy protection, crowdsensing coalition, data analytics, privacy-preserving crowdsensing.
\end{IEEEkeywords}

\section{Introduction}

The proliferation of mobile devices with built-in sensors has made mobile crowdsensing an efficient sensing paradigm especially in people-centric and Internet of Things (IoT) services. Crowdsensing users collect sensing data using their personal mobile devices, e.g.,~mobile phones and IoT gadgets. However, the development of crowdsensing services is impeded by many challenges, especially the criticism on the privacy protection of crowdsensing users. Service providers require true data which is a key factor in optimizing data originated services~\cite{vaidya2006privacy}. This introduces contradicting incentives of maximizing the privacy protection of users and the prediction accuracy of service providers. Most of the existing incentive models in the literature are monetary motivated with sole profit maximization objective, e.g.,~\cite{YangXueFangEtAl2012,XuXiangYang2015,luo2016incentive}, while the privacy incentive of users is neglected. Therefore, conventional monetary-based incentive models are inapplicable in privacy preserving crowdsensing systems, and new privacy-aware incentive models are required. Several major questions related to developing privacy-aware incentive models in mobile crowdsensing arise. First, how does the crowdsensing service define the contributions and payoff allocations of users with varying privacy levels? Second, do crowdsensing coalitions change the attained privacy of the cooperative users? Third, how do cooperative users divide the coalition payoff among themselves?

This article provides answers for the aforementioned questions by presenting a novel incentive framework for privacy preservation and accuracy maximization in mobile crowdsensing. The sensing users select their preferred data anonymization levels without knowing the privacy preferences of the other users. The data anonymization is inversely proportional to the accuracy of data analytics of the service provider. Accordingly, the users are paid based on their marginal contributions to the service accuracy. The users can be also penalized with a negative payoff if they cause a marginal harm to the service accuracy, e.g.,~an outlier providing misleading data. Moreover, a set of $k$~cooperative users can jointly work by forming a crowdsensing coalition, increasing the anonymity privacy protection measured by the $k$-anonymity metric. The total coalition payoff is then divided among the cooperative users based on their marginal contributions to the coalition's data quality. Our experiments on a real-world dataset of crowdsensing activity recognition system show that the payoff allocation of a particular user does not directly depend on the contributed data size but on the data quality. Likewise, the payoff allocation is found to decrease as the privacy level increases.

The rest of this article is organized as follows. We first present an overview of mobile crowdsensing in people-centric and IoT services and review some related incentive mechanisms. Next, we discuss the privacy preservation in mobile crowdsensing. Then, we propose an incentive framework for privacy preservation and accuracy maximization in crowdsensing services. After that, we present numerical experiments based on a real-world crowdsensing dataset. Finally, we outline some interesting research directions and conclude the article.

\section{Mobile Crowdsensing and IoT}

This section first gives an overview of mobile crowdsensing in IoT and then reviews some monetary incentive mechanisms in mobile crowdsensing. 

\subsection{Architectures and Data Management}

In mobile crowdsensing, mobile devices and human intelligence are jointly adopted for collecting sensing data regardless of geographic separation among users and service providers. As shown in Figure~\ref{fig:sensing_architecture}, the design of mobile crowdsensing services includes the following stages:

\begin{figure*}
\begin{centering}
\includegraphics[width=1\linewidth,trim= 0 1cm 0 0]{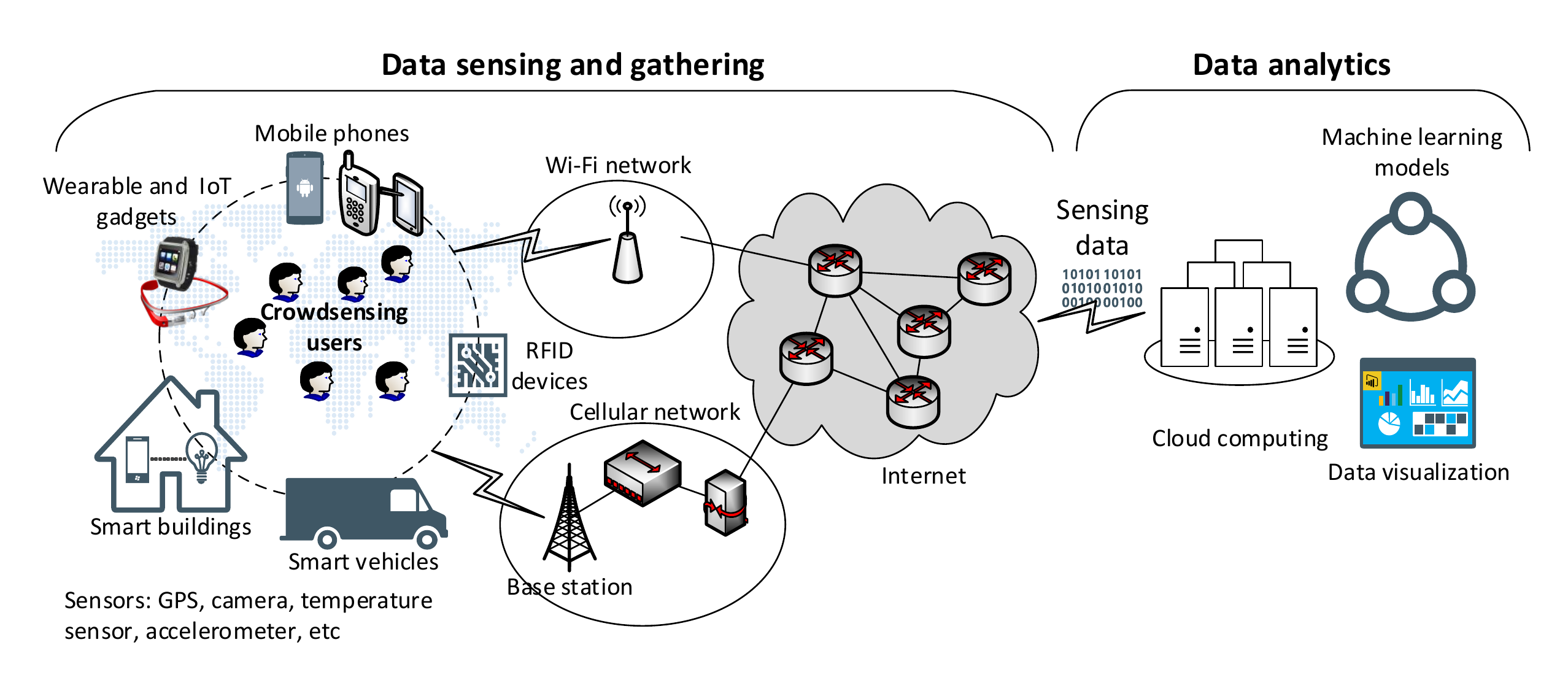}
\par\end{centering}
\caption{System model of mobile crowdsensing.\label{fig:sensing_architecture}}
\end{figure*}

\begin{itemize}
\item \emph{Data Sensing and Gathering}: Crowdsensing users sense and collect data using mobile devices including phones, wearable devices, and in-vehicle sensing devices. Users can also annotate the sensory data with subjective observations and reports such as their emotions and surrounding events. The data is sent to the cloud server through various types of networks including cellular and Wi-Fi networks.
\item \emph{Data Analytics}: After receiving the raw data from the users, cloud computing can be used to store and process the large-scale data. Data analytics, e.g.,~machine learning methods, are typically applied to extract useful information and make effective predictions. Services also support data visualization, generate reports, and provide platforms to share the outcomes with other collaborative entities, e.g.,~social networking services.
\end{itemize}

\subsection{Applications}

Mobile crowdsensing has become an efficient sensing paradigm in people-centric and IoT services. People-centric services contain sensing, computing, and communication components that aim to assist human life. The following are some pertinent crowdsensing applications.
\begin{itemize}
\item \emph{Traffic Monitoring}: Mobile Millennium\footnote{\url{http://traffic.berkeley.edu}, accessed on 18 December 2016.} is a traffic crowdsensing service. Millennium collects geolocation data from taxi drivers. It also assimilates other data obtained in realtime from radars, loop detectors and historical databases. The traffic information can be accessed by drivers for accurate real-time traffic conditions, e.g.,~traffic congestion points.
\item \emph{Wi-Fi Sharing}: WiFi-Scout\footnote{\url{http://wifi-scout.sns-i2r.org}, accessed on 18 December 2016.} is a crowdsensing service for sharing reviews and connection quality of Wi-Fi hotspots. Users can easily search for free and paid Wi-Fi hotspots covering the locations that they will be visiting. The users are also rewarded based on their compliance and review quality.
\item \emph{Healthcare}: PatientsLikeMe\footnote{\url{https://www.patientslikeme.com}, accessed on 18 December 2016.} is a healthcare crowdsensing service that collects health data from patients. The patients provide their experience on medication, supplement, or devices. PatientsLikeMe also sells the collected data to pharmaceutical companies in order to improve and develop effective medication and healthcare equipment.
\end{itemize}

\subsection{Monetary Incentive Models}

Mobile crowdsensing should incorporate efficient incentive mechanisms to attract and retain enough crowdsensing users. In~\cite{musthag2011exploring}, the authors compared the resulting data quality and user compliance of three incentive schemes. The \emph{uniform} scheme pays user at a fixed rate of $4$ cents per completed task. The \emph{variable} scheme selects the payoff in the range of $2$ to $12$ cents based on the required task and user performance. Finally, the \emph{hidden} scheme includes a lottery factor in defining the payoff values where the users are not informed of the expected payoff before completing the task. The study showed that the variable scheme reduces the total cost by $50\%$ compared to the uniform scheme for the same completion rate and performance. The hidden scheme is found to be the least effective incentive scheme.

\begin{figure}
\begin{centering}
\includegraphics[width=1\columnwidth,trim= 1.5cm 0.5cm 1cm 0]{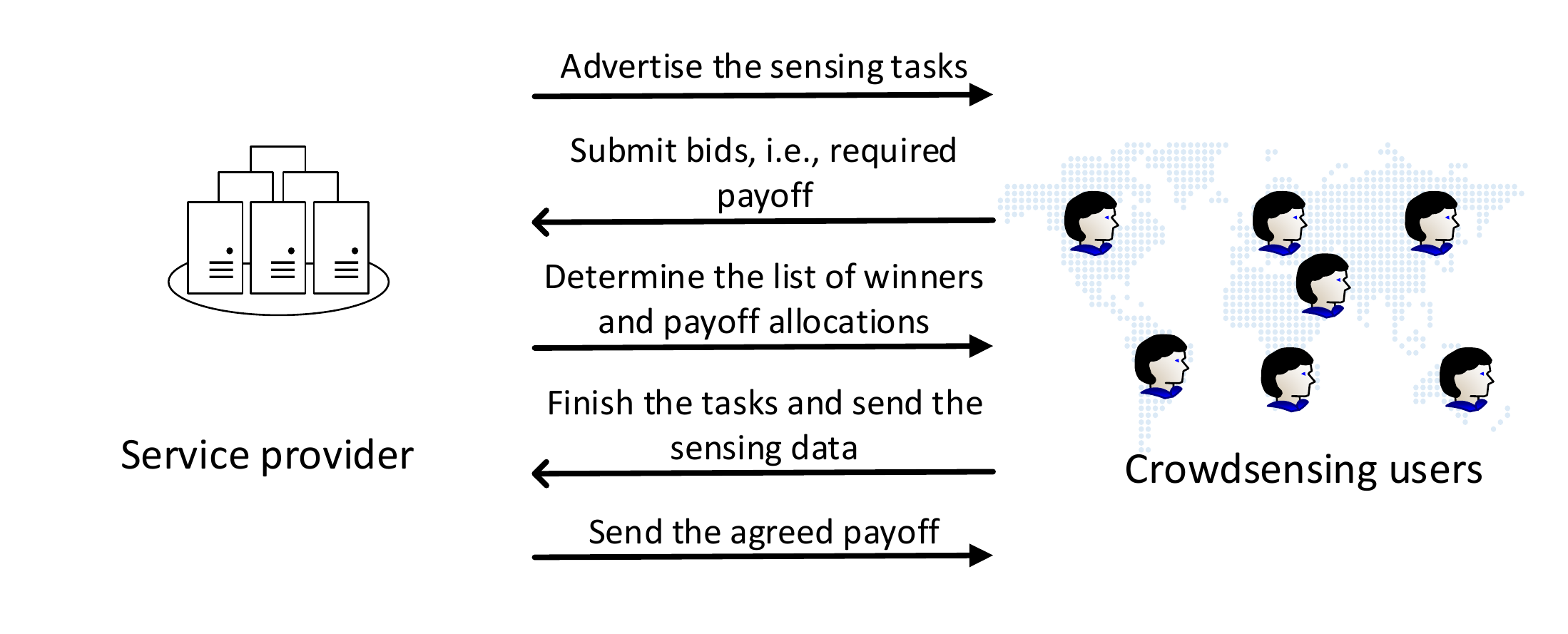}
\par\end{centering}
\caption{Crowdsensing incentive mechanism as a reverse auction.\label{fig:reverse_auction}}
\end{figure}

Next, we review monetary incentive mechanisms for mobile crowdsensing with an emphasis on reverse auction mechanisms~\cite{Klemperer2004} as they fit well and are commonly applied for mobile crowdsensing with multiple users. As shown in Figure~\ref{fig:reverse_auction}, a typical reverse auction framework occurs between the crowdsensing users and service. The crowdsensing users compete among themselves to perform the sensing task. The service provider first announces the description of the crowdsensing tasks to potential mobile users. Users are rational entities and will set their bids based on the cost of the crowdsensing task. In order to maximize the utility of the crowdsensing service, the auction system determines the task assignment and payoff of each user including both selected and rejected bids. For example, the crowdsensing tasks are assigned to the winning users with the lowest bids to perform the crowdsensing tasks and submit the data to the service. The service provider will provide the agreed payoff to the winning users. Table~\ref{tab:summary_auctions} provides a summary of the monetary incentive models reviewed in this section. From the table, ``risk-neutral'' means that the user is unaware of the loss of its payoff, e.g.,~when choosing between guaranteed \$$5$ and conditioned \$$10$ payoffs. A ``profitable'' solution guarantees a nonnegative utility for the service provider. An ``individual rational'' solution guarantees a nonnegative utility for each user. A ``truthful'' solution guarantees that the users cannot increase their payoff by submitting misleading bids for the crowdsensing task. Therefore, a truthful incentive mechanism provides a dominant strategy for rational users in bidding their true cost of performing the crowdsensing task.

\begin{table*}
\caption{Summary of the monetary incentive models in mobile crowdsensing.\label{tab:summary_auctions}}
\centering{}%
\begin{tabular}{>{\raggedright}p{2.3cm}>{\raggedright}p{3cm}>{\raggedright}p{3cm}>{\raggedright}p{3cm}>{\raggedright}p{3cm}}
\hline 
\textbf{\noun{\small{}model}} & \textbf{\noun{\small{}Main entities}} & \textbf{\noun{\small{}Payoff scheme }} & \textbf{\noun{\small{}Maximization objective}} & \textbf{\noun{\small{}Solution Properties}}\tabularnewline
\hline 
\hline 
{\small{}Bayesian auction~\cite{CaoBrahmaVarshney2015}} & {\small{}Multiple risk-neutral users } & {\small{}Threshold winner payoff} & {\small{}The target tracking accuracy} & {\small{}Bayesian Nash equilibrium (profitable and individual rational)}\tabularnewline
\hline 
{\small{}Sealed-bid auction~\cite{YangXueFangEtAl2012}} & {\small{}Fixed budget with risk-neutral users} & {\small{}Threshold winner payoff} & {\small{}The service utility (more user and less payoff)} & {\small{}Profitable and individual rational}\tabularnewline
\hline 
{\small{}Stackelberg competition~\cite{YangXueFangEtAl2012}} & {\small{}A leader (service) and followers (users) } & {\small{}Threshold winner payoff} & {\small{}The service utility} & {\small{}Nash equilibrium (profitable and individual rational)}\tabularnewline
\hline 
{\small{}Vickrey auction~\cite{XuXiangYang2015}} & {\small{}Multiple risk-neutral users} & {\small{}Contribution-dependent payoff} & {\small{}Data integrity} & {\small{}Profitable and truthful}\tabularnewline
\hline 
{\small{}All-pay auction~\cite{luo2016incentive}} & {\small{}Risk-averse and risk-neutral users} & {\small{}All-pay contribution-dependent payoff} & {\small{}The service utility} & {\small{}Nash equilibrium (profitable and individual rational)}\tabularnewline
\hline 
\end{tabular}
\end{table*}

We divide the incentive schemes into two main categories of \emph{threshold winner} and \emph{contribution-dependent} payoffs. 

\subsubsection{Threshold Winner Payoff}

In this payoff scheme, only the winning users will be paid for performing the sensing task and there is no payoff allocation for rejected users. For example, the authors in~\cite{CaoBrahmaVarshney2015} presented a Bayesian reverse auction model for target tracking with crowdsourcing, assuming that the value estimate of a user can be drawn from a continuous probability distribution. The residual energy of the mobile devices has an impact on the prior distribution of the user bids. The objective of this model is maximizing the total target tracking utility of the service by solving the multiple-choice knapsack problem. Likewise, the authors in~\cite{YangXueFangEtAl2012} proposed two complementary payoff scenarios of \emph{user-centric} and \emph{platform-centric} schemes. In the user-centric scheme, the service defines the payoff using a reverse auction by following the steps shown in Figure~\ref{fig:reverse_auction}. In the platform-centric scheme, the crowdsensing problem is formulated as a Stackelberg game. The Nash equilibrium is solved using backward induction and found to be unique. A major limitation of~\cite{CaoBrahmaVarshney2015,YangXueFangEtAl2012} is assuming a known prior distribution of user bids. In the real world, users can collude and submit misleading bids to increase their own payoff. This problem is solved in contribution-dependent payoff schemes as discussed next.

\subsubsection{Contribution-Dependent Payoff}

A practical incentive mechanism requires all participants to be truthful. One principal way in achieving truthful user interaction is by choosing an appropriate pricing scheme where the payoff  allocations of participants are not solely defined by their bids. The authors in~\cite{XuXiangYang2015} applied the Vickrey-Clarke-Groves (VCG) reverse auction with the objective of minimizing the sum of payoff values to crowdsensing users. A user is paid based on the difference between the sum of costs with and without that particular user. Reporting truthful bids is a weakly-dominant strategy in the VCG auction. The authors in~\cite{luo2016incentive} modeled the mobile crowdsensing problem as an all-pay auction where the crowdsensing users are not required to submit their bids at the beginning of the auction. Instead, the payoff is calculated based on the user contributions after completing the sensing tasks. The users with the highest contribution receive a payoff while the rest of the users do not receive any payoff allocation.

\section{Privacy Preservation in Mobile Crowdsensing}

Even though most of the existing works in the literature focus on monetary incentive models to achieve the maximum possible payoff allocation, privacy preservation is still a top priority for crowdsensing users. In this section, we first discuss the data anonymization properties which can be used to measure the privacy protection. Then, we discuss the challenges of privacy preservation in mobile crowdsensing.

\subsection{Privacy Properties and Data Anonymization}

Mobile crowdsensing comes with challenging privacy issues. In particular, crowdsensing users are typically concerned that their personal information can be leaked from the collected data. Personal information of users can be categorized into three main classes:
\begin{itemize}
\item \emph{Explicit identifiers} are the data attributes which directly reveal the user identity, e.g.,~full name and social security number.
\item \emph{Non-explicit identifiers} can be combined with background knowledge to reveal the user identity, e.g.,~zip code and birth date.
\item \emph{Sensitive attributes} can be utilized to extract private information about the user, e.g.,~realtime activity tracking using accelerometer data~\cite{kwapisz2011activity}.
\end{itemize}
Explicit identifiers should be completely removed before trading the crowdsensing data among businesses. To protect the non-explicit identifiers and sensitive attributes, data anonymization methods can be applied to sensing data.

Privacy is defined by the information gain of an adversary. The following syntactic privacy properties can be used to define of the privacy protection requirements.
\begin{itemize}
\item \emph{$k$-anonymity}~\cite{Sweeney2002}: This property is developed to guarantee that a data sample of a particular user in public datasets cannot be re-identified by potential intruders. Specifically, for a crowdsensing service to possess the $k$-anonymity property, each user should not be distinguishable from at least $k-1$ other users. For example, a user should be unidentifiable by combining the available gender and birth date crowdsensing data. This can be achieved by transformation techniques, such as identity generalization and suppression, to reduce the granularity of the data. For example, the birth dates can be replaced by date ranges instead of the exact values.
\item \emph{$l$-diversity}~\cite{machanavajjhala2007diversity}: The $k$-anonymity does not work well if the sensitive data attributes lack diversity. For example, if a few users of a healthcare crowdsensing service used a particular zip code and are infected by a disease, then background knowledge can be used to reveal the health privacy of a user which is known by the adversary to use that zip code. In order to avoid this privacy threat, the $l$-diversity property requires that each equivalence class has at least $l$ ``well-represented'' values. An equivalence class is a set of data samples with the same anonymized data attributes.
\item \emph{$t$-closeness}~\cite{LiLiVenkatasubramanian2007}: The $t$-closeness property requires the distribution of sensitive values within each equivalence class to be \textquotedblleft close\textquotedblright{} to their distribution in the entire original dataset. $t$-closeness is an extension of the $l$-diversity model as it takes the distribution of sensitive values into account. $t$-closeness can be achieved by adding random noise to sensitive data attributes. For example, adding Gaussian noise to accelerometer data can restrict the tracking of particular activities.
\end{itemize}

\subsection{Challenges of Privacy Preservation in Mobile Crowdsensing}

The authors in~\cite{pournajaf2016participant} reviewed the privacy threats and protection methods during the task management in mobile crowdsensing. A taxonomy of privacy methods was provided including pseudonyms, connection anonymization, and spatial cloaking. The authors also highlighted the challenging process of defining the user contribution in incentive-based task assignment. The authors in~\cite{ganti2011mobile} discussed the privacy and data integrity of mobile crowdsensing. The privacy is observed to be user-dependent.

Achieving the syntactic privacy properties can reduce the accuracy of data analytics algorithms. Applying a strict data anonymization to all users results in a poor accuracy of the data analytics. Instead, the users can be given the choice of setting their preferable data anonymization level such that reliable users receive high payoff allocation. The trade-off between the privacy preservation and accuracy maximization should also be taken into consideration which is the main objective of the next section.

\section{Incentive Mechanism for Privacy Preserving Crowdsensing}

In this section, we introduce a privacy preserving incentive framework for mobile crowdsensing where participating users can protect their private data by data anonymization. The level of data protection will accordingly be used to set the resulting payoff allocation such that the users have an incentive in providing their true data. We first present the system model and major entities. Then, we discuss the proposed incentive framework which is intended to maximize the accuracy of data analytics while preserving the privacy of the crowdsensing users.

\subsection{System Model}

\begin{figure}
\begin{centering}
\includegraphics[width=1\columnwidth,trim= 1cm 0.5cm 0.5cm 0]{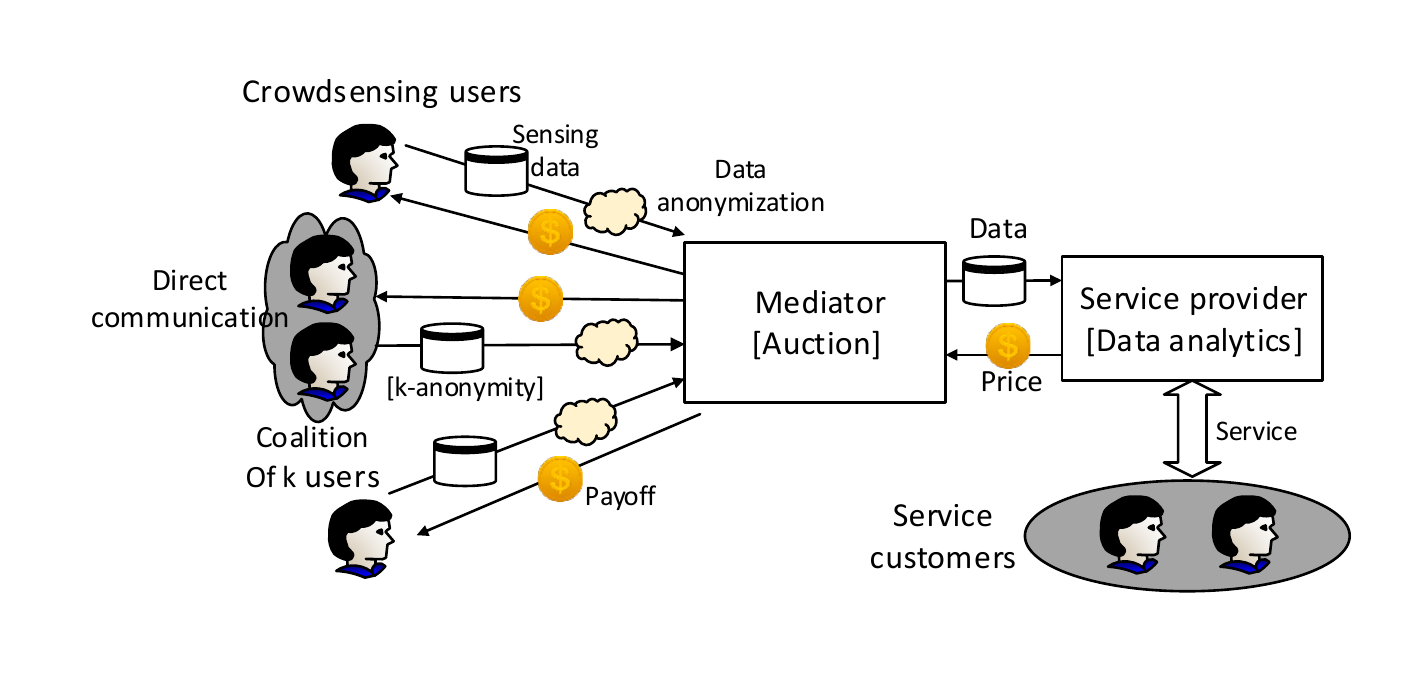}
\par\end{centering}
\caption{System model of the privacy preserving crowdsensing framework supporting both data anonymization and identity generalization through crowdsensing coalition formulation. Cooperative users are connected using device-to-device (D2D) communication.\label{fig:system_model}}
\end{figure}

As shown in Figure~\ref{fig:system_model}, the crowdsensing system under consideration consists of the following three main entities:
\begin{itemize}
\item Crowdsensing\textbf{ users} are the participants which collect sensing data using their personal mobile devices, e.g.,~mobile phones and IoT gadgets. The contribution of a particular user to the crowdsensing community is defined based on the quality of the sensing data from the data analytics perspectives. A user with positive contribution to the sensing process is considered \emph{pivotal}. The users can apply data anonymization, e.g.,~adding noise to the sensing data, to protect their privacy and personally-identifying information. Additionally, crowdsensing coalitions can be built as an efficient scheme for achieving $k$-anonymity protection, where $k$ is the number of cooperative users in the coalition.
\item A \textbf{service provider} buys data from the crowdsensing users through a mediator, applies data analytics, and delivers a service to a set of customers. The provider makes a profit by charging the customers a subscription fee.
\item A \textbf{mediator} is the auction management entity that controls the exchange of data between the crowdsensing users and the service provider. Moreover, the mediator divides the payoff received from the service provider among the crowdsensing users based on their contributions to the crowdsensing system.
\end{itemize}
We next discuss the privacy preserving model through which the crowdsensing users can sell data to the service provider and receive a payoff according to their individual contributions as illustrated in Figure~\ref{fig:system_model}. First, we define the individual contributions and resulting payoffs of the users from data analytics perspectives. Second, we develop a privacy preserving mechanism which gives the users an incentive for contributing their true data with the least possible data anonymization level. Third, we consider the case where the users can form a crowdsensing coalition for identity generalization, and we present a fair payoff allocation among the cooperative users.

\subsection{Data Analytics}

Crowdsensing data $\mathcal{D}=\left\{ \left(\mathbf{x}_{i},y_{i}\right)\right\} _{i=1}^{L}$ usually includes tuples of sensing feature set $\mathbf{x}_{i}\in\mathbb{R}^{M}$ and a class label $y_{i}\in\mathbb{R}$, where $L$ is the number of data tuples and $M$ is the number of data attributes. The feature set $\mathbf{x}_{i}$ includes the sensing data, e.g.,~images in vision services and geographic coordinates in transport services. The class label $y_{i}$ contains human input and is only available in supervised data analytics, e.g.,~specifying accident events in transport services. After collecting sufficient data, the service provider applies data analytics methods, e.g.,~deep learning~\cite{lecun2015deep}, to build data originated services. For example, transport services can provide accurate prediction of vehicle arrival times and road congestions. We denote the accuracy function of the data analytics model trained using dataset $\mathcal{D}$ as $f\left(\mathcal{D}\right)$. $f\left(\mathcal{D}\right)$ measures the performance of the service in providing accurate prediction of the ground truth.

\subsection{Incentive Mechanism Design}

We consider a set of $N$ users which are connected to a privacy preserving crowdsensing service. Each user $n$ generates true sensing data $\mathcal{D}_{n}$ and selects a data anonymization level $p_{n}$. The data anonymization can be performed by adding random noise to the true data $\mathbf{x}_{i}$ subject to $p_{n}$, e.g.,~Gaussian noise $\mathcal{N}\left(0,p_{n}\mathbf{I}_{M}\right)$ with zero mean and a variance of $p_{n}$, where $\mathbf{I}_{M}$ is the identity matrix of size $M$. Each user submits its anonymized data $\tilde{\mathcal{D}}_{n}$ and data anonymization level $p_{n}$ to the crowdsensing mediator, without knowing the preferences of the other users. The full anonymized dataset $\tilde{\mathcal{D}}=\underset{1\leq n\leq N}{\cup}\tilde{\mathcal{D}}_{n}$ and data anonymization preferences $\mathcal{P}=\left\{ p_{1},\ldots,p_{N}\right\} $ are collected by the mediator from all users. According to the VCG auction~\cite{Klemperer2004}, the mediator calculates the payoff of user~$n$ as follows:
\begin{equation}
F_{n}=f(\tilde{\mathcal{D}})-f(\mathcal{\tilde{\mathcal{D}}}_{-n}),\label{eq:utility_1}
\end{equation}
where $\mathcal{\tilde{\mathcal{D}}}_{-n}$ is the anonymized data after excluding the data of user $n$. The following three cases for the payoff function exist:
\begin{itemize}
\item If $F_{n}>0$, the user will receive a positive payoff allocation of $F_{n}$ as its data contribution increases the accuracy. These users are called pivotal.
\item If $F_{n}=0$, the user does not change the crowdsensing choice nor the service accuracy. Such users receive zero payoff and can be advised to decrease their data anonymization level.
\item If $F_{n}<0$, the user has a negative contribution, e.g.,~excessive data anonymization, and will accordingly be penalized with a negative payoff. The data collected from such users should not be used in the data analytics.
\end{itemize}
Sending the true data to the service provider is a weakly-dominant strategy under the VCG rules regardless of the data anonymization levels of the other users.

\subsection{Crowdsensing Coalition}

A set of $k$ users can cooperate to form a crowdsensing coalition, denoted by $\mathcal{K}$, which increases the privacy level by providing the data of the cooperative users under one generalization identity and achieving $k$-anonymity privacy protection. Those $k$ users must be connected using device-to-device~(D2D) communication without traversing the service provider. The generalization identity guarantees that a data sample cannot be used to identify its source from the $k$ cooperative users. $\mathcal{K}$ is a virtual alliance of users which work collectively and are seen as one sensing entity by the service provider. Specifically, the service provider cannot identify the source of data samples as a particular data sample can relate to any of the $k$ cooperative users. The payoff of the coalition is
\begin{equation}
F_{\mathcal{K}}=f(\tilde{\mathcal{D}})-f(\mathcal{\tilde{\mathcal{D}}}_{-\mathcal{K}}),\label{eq:utility_2}
\end{equation}
where $\mathcal{\tilde{\mathcal{D}}}_{-\mathcal{K}}$ is the anonymized data after excluding the data from all users in the coalition $\mathcal{K}$. Solution concepts from cooperative game theory, such as the Shapley value and Nash bargaining solution~\cite{peleg2007introduction}, can be applied to share the resulting payoffs among the cooperative users in the coalition $\mathcal{K}$. From the Shapley value, the payoff allocation, i.e.,~monetary payment, of each user is defined based on its contribution to the coalition.

\section{Numerical Results}

In this section, we present numerical experiments to evaluate the performance of the proposed privacy preserving framework.

\subsection{System Setup}

In this section, we use a real-world dataset~\cite{kwapisz2011activity} of crowdsensing activity recognition system of six activities including walking, jogging, upstairs, downstairs, sitting, and standing. The dataset includes $L=1,098,207$ samples of accelerometer data which were collected by $N=36$~users. The mobile devices sampled at a rate of $20$Hz resulting in $M=120$ data features of framed $3$-axial acceleration. We assume that the service provider uses deep learning~\cite{lecun2015deep} to develop the prediction service. The service provider buys the crowdsensing data from the users through the auction mediator and sells an activity tracking service to customers.

We assume that Users~2 and 3 protect their sensitive activities by adding varied levels of Gaussian noise $\mathcal{N}\left(0,p_{n}\mathbf{I}_{M}\right)$ to the acceleration data. Accordingly, Users~2 and 3 acquire the $t$-closeness property, where $t$ is equal to the variance of the added noise $p_{n}$. The payoff of each user is defined based on the payoff rule in~(\ref{eq:utility_1}). Moreover, Users~2 and 3 can collaborate in the crowdsensing coalition $\mathcal{K}$ to acquire the $k$-anonymity protection, where $k=2$ for two cooperative users. The coalition's total payoff is defined based on the payoff rule in~(\ref{eq:utility_2}), while the payoff sharing among Users~2 and 3 is defined according to the Shapley value.

\subsection{User Contributions and Pivotal Users}

\begin{figure}
\begin{centering}
\includegraphics[width=1\columnwidth]{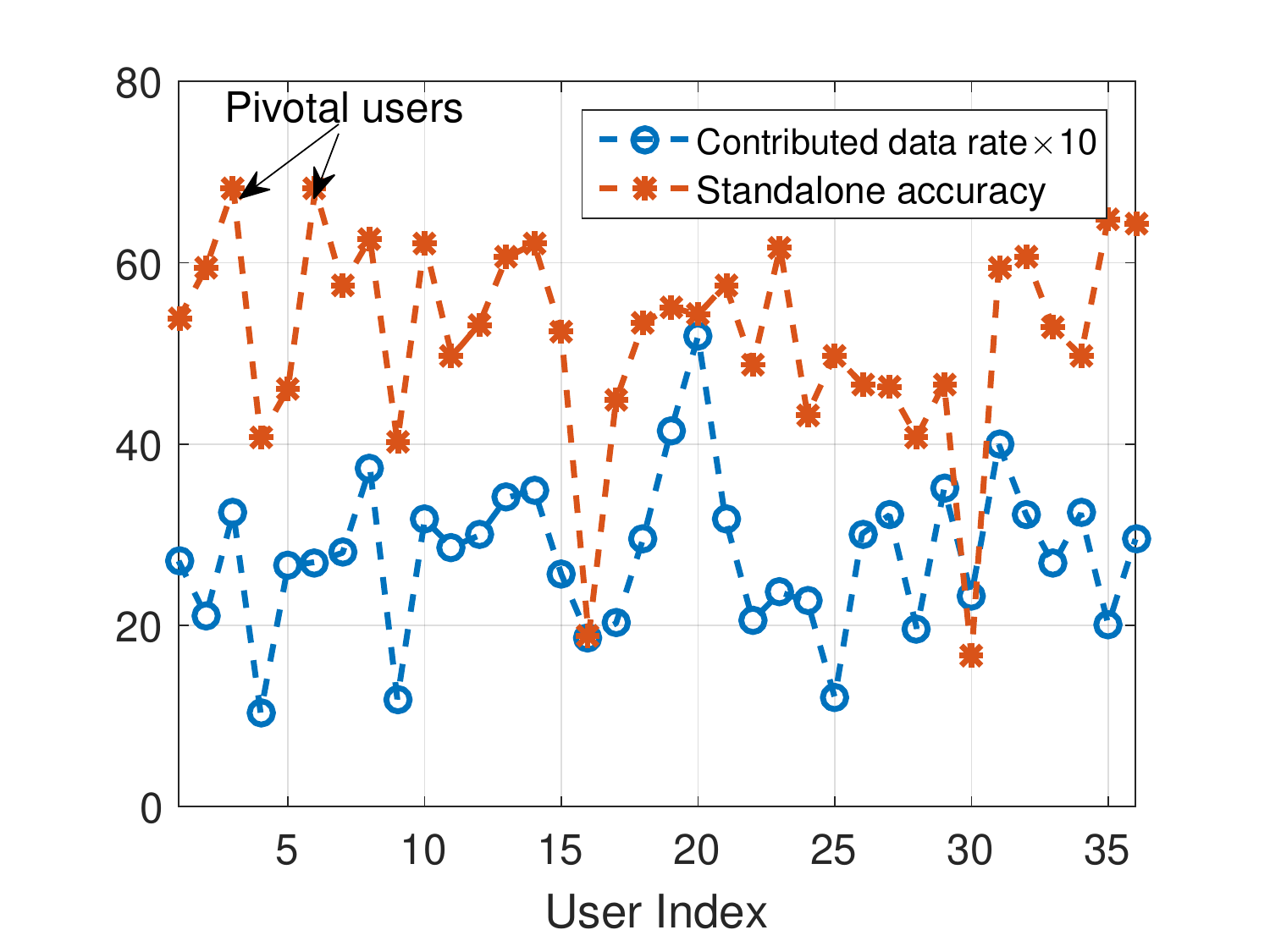}
\par\end{centering}
\caption{User contribution to the crowdsensing service.\label{fig:agent_contribution}}
\end{figure}

Figure~\ref{fig:agent_contribution} shows the contributed data rates from each user and the resulting service accuracy $f(\cdot)$ by training a deep learning model on the data of each user separately. Two key results can be noted. Firstly, the data rate varies among different users. However, there is no correlation between the service accuracy from the data analytics perspective and the contributed data rate from the sensing perspective. The service accuracy depends on the quality of the used mobile device, user's performance during task execution, and data annotation. For example, User~1 contributes more data than that of User~2, while the accuracy resulting from the data of User~1 is lower than that of User~2. Secondly, Users~3 and 6 are pivotal and they score the highest standalone accuracy values of $68.3\%$ and $68.1\%$, respectively. The standalone accuracy for the rest of the users is less than $64\%$. The pivotal users are important to the service provider to ensure high service accuracy.

\subsection{The Impact of Privacy on Accuracy}

\begin{figure}
\begin{centering}
\subfloat[\label{fig:privacy_accuracy}]{\begin{centering}
\includegraphics[width=1\columnwidth]{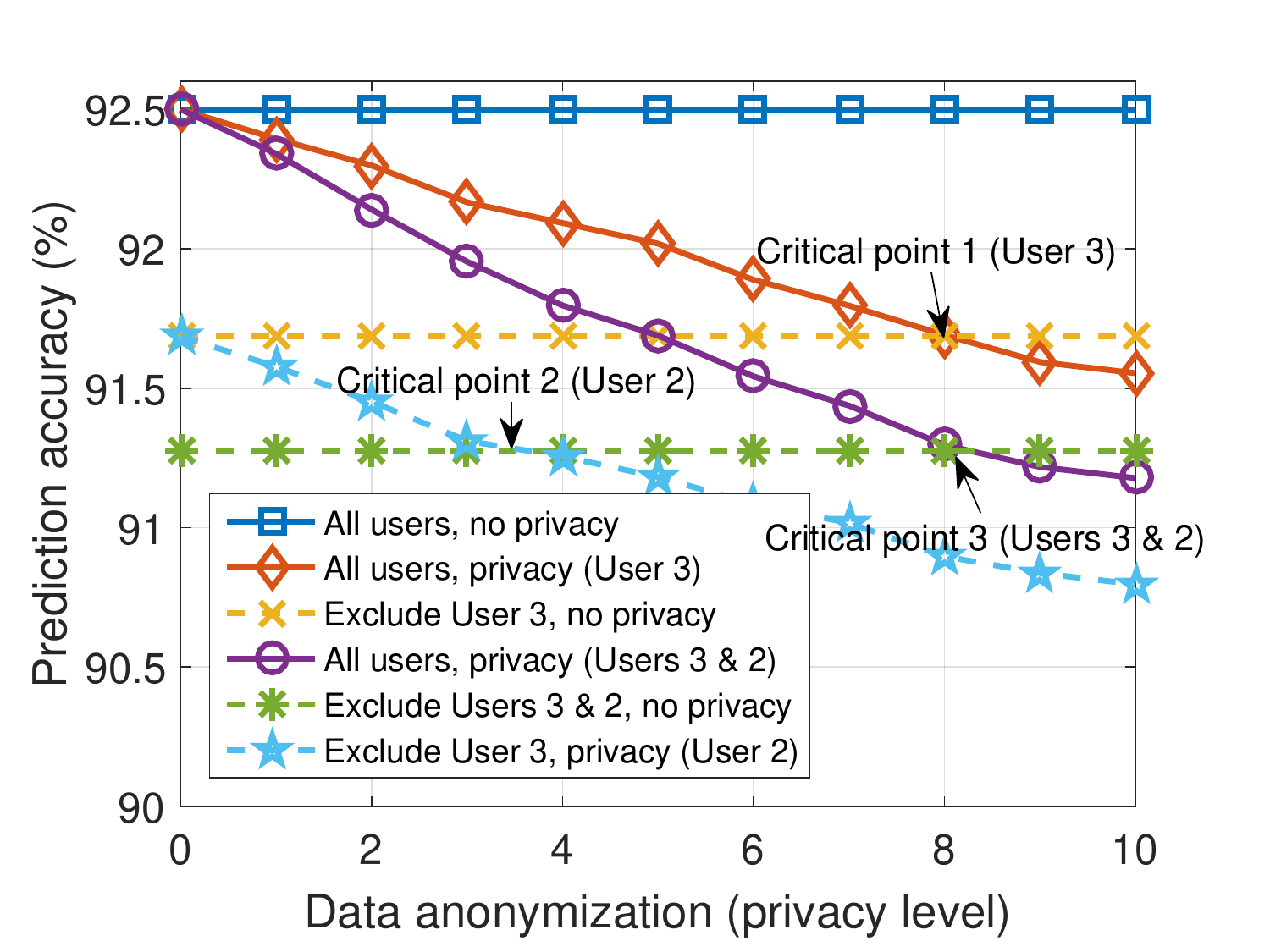}
\par\end{centering}

}
\par\end{centering}

\begin{centering}
\subfloat[\label{fig:privacy_payoffs}]{\begin{centering}
\includegraphics[width=1\columnwidth]{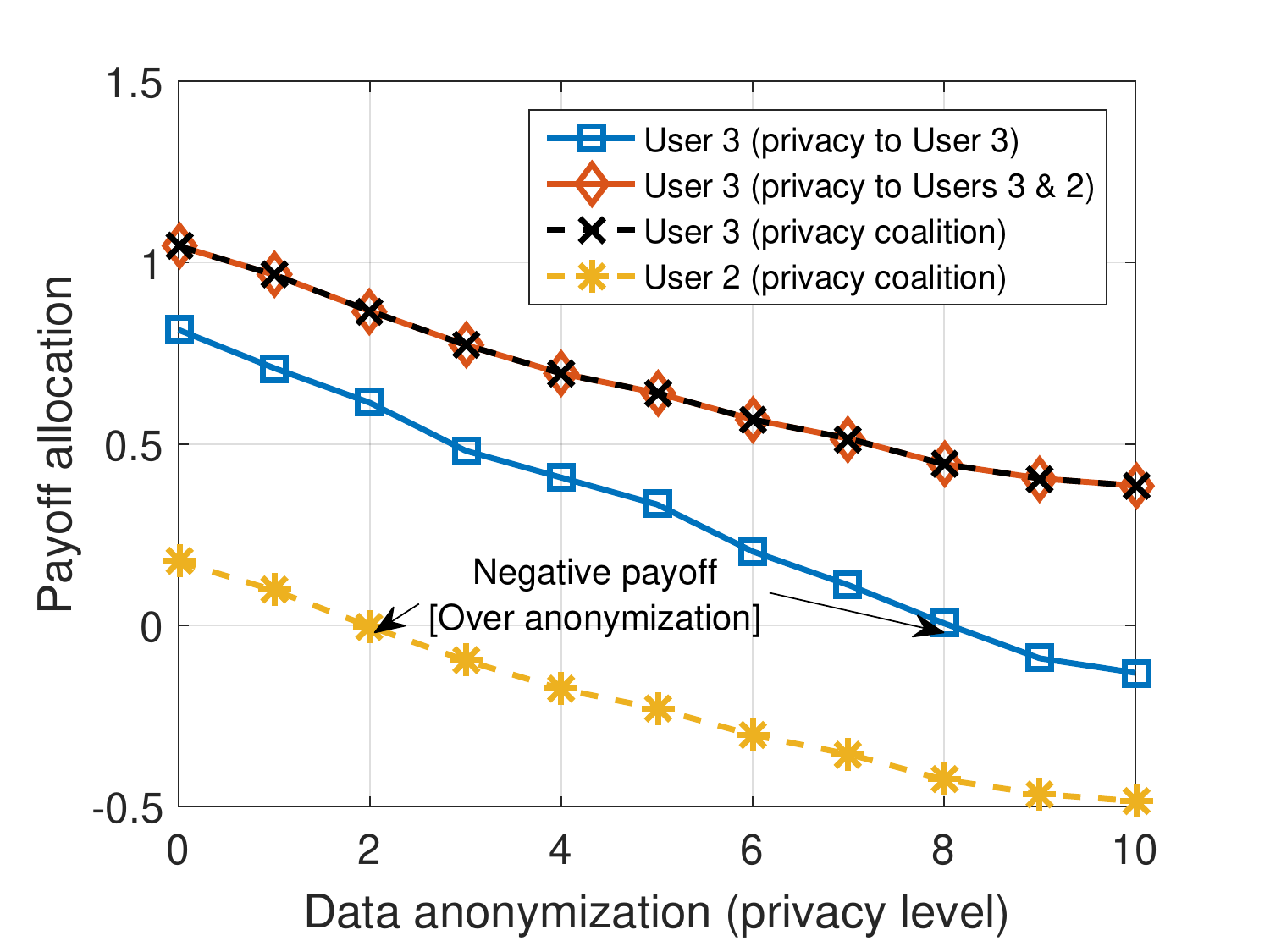}
\par\end{centering}

}
\par\end{centering}

\caption{Performance of the proposed privacy preserving framework under varied privacy levels. (a)~The resulting accuracy of the deep learning service trained on the crowdsensing data. (b)~The payoff allocation of Users~2 and~3. The privacy level is equal to the variance of the added Gaussian noise.}
\end{figure}

In Figure~\ref{fig:privacy_accuracy}, we consider the impact of the data anonymization level on the accuracy of the crowdsensing service. Several important results are observed. Firstly, there is an inverse relationship between the prediction accuracy and the data anonymization level. The maximum service accuracy of $f(\mathcal{D})=92.5\%$ is achieved when all users provide true data samples without any anonymization. This maximum value decreases as User~3 increases the level of data anonymization. High level of data anonymization can be required by the users to protect their privacy. Secondly, the service provider has an incentive of rejecting users with high data anonymization levels. For example, the service will reject User~3 when its data anonymization level is greater than $8$ which is labeled as ``critical point 1'' in Figure~\ref{fig:privacy_accuracy}. This is due to the resulting harm to the overall system accuracy. Thirdly, the prediction accuracy decreases as more users adopt the data anonymization scheme. For example, the accuracy is negatively affected when both Users~2 and 3 apply the data anonymization compared to the case of User~3 only. Accordingly, the crowdsensing system has an incentive for reducing the number of users applying the data anonymization scheme. As presented next, this can be achieved by increasing the payoff allocation of users which provide their true data.

\subsection{Payoff Allocation}

Figure~\ref{fig:privacy_payoffs} shows the payoff allocation of Users~2 and 3 under the varied data anonymization levels. Firstly, the payoff allocation of any user decreases as its data anonymization level increases. For high data anonymization level which is equal to or greater than the over anonymization levels specified in Figure~\ref{fig:privacy_payoffs}, the users will be penalized by receiving negative payoff. Secondly, pivotal users receive a higher payoff compared to normal and low-performing users, e.g.,~the payoff of User~3 is greater than that of User~2. For the crowdsensing coalition case, the payoff allocation to the cooperative users is found using the Shapley value which reflects the individual contribution of each user. The cooperative users receive not only the same payoff in both the crowdsensing coalition and the standalone cases, but also a higher level of the $k$-anonymity privacy protection.

\section{Future Directions}

Based on the proposed incentive framework, the following open research directions can be further pursued. 

\subsection{Cooperation and Competition Among Service Providers}

To collect high-quality data, service providers may cooperate or compete with each other to attract and retain crowdsensing users. With cooperation, service providers collude to set payoff strategies which maximize their profit as a cooperative coalition. In the competitive scenario, service providers can apply non-cooperative game and Nash equilibrium solutions for the service's subscription fee and crowdsensing data's prices. The strategic interaction among providers can also benefit the users in making higher revenues.

\subsection{Incentive Mechanism Design for Fog Computing}

Analyzing the crowdsensing data can be computationally expensive. Fog computing provides a solution by allowing partial data processing at the mobile devices owned by users. In such a design, the users are paid not only for the sensing data, but also for the available computing power. Incentive mechanisms are required to attract large contributions from users as fog nodes. Likewise, mobile devices come with varying hardware resources; methods for defining the user contributions in fog computing are also required.

\subsection{Dynamic and Heterogeneous Crowdsensing}

Crowdsensing users can be heterogeneous in term of the sensing precision and technical experience. Thus, the service provider has an incentive of attracting powerful users by increasing their payoff allocations, and the incentive mechanism has to optimize these payoff values. Additionally, users asynchronously join and leave the crowdsensing system. Stochastic optimization methods, e.g.,~Markov decision processes, can be formulated to determine the optimal payoff rates over time, e.g.,~to attract users during the off-peak times.

\section{Conclusion}

Privacy awareness has the potential of significantly boosting the performance of mobile crowdsensing, attracting more sensing users, and enabling the protection of privileged information. This article has presented an incentive mechanism for privacy preservation and accuracy maximization in mobile crowdsensing. It has been shown that the coalition strategy can be used by users to send their data under one generalized identity, increase the $k$-anonymity privacy protection, and share the resulting payoffs among cooperative users based on their individual sensing contribution. The proposed incentive framework has been evaluated using a real-world crowdsensing dataset. Finally, open research directions have been presented.

\section*{Acknowledgment}

This work was supported in part by Singapore MOE Tier 1 (RG18/13 and RG33/12) and MOE Tier 2 (MOE2014-T2-2-015 ARC4/15 and MOE2013-T2-2-070 ARC16/14).

\bibliographystyle{ieeetr}
\bibliography{references}

\section*{Biographies}
\begin{IEEEbiographynophoto}
{Mohammad Abu Alsheikh}
[S'14] (mohammad027@e.ntu.edu.sg) received the B.Eng. degree in computer systems engineering from Birzeit University, Palestine, in 2011. Between 2010 and 2012, he was a Software Engineer working on developing robust web services, Ajax-based web components, and smartphone applications. He is currently a Ph.D. candidate in the School of Computer Science and Engineering, Nanyang Technological University, Singapore. His research interests include machine learning in big data analytics, mobile sensing technologies, and sensor-based activity recognition.
\end{IEEEbiographynophoto}

\begin{IEEEbiographynophoto}
{Yutao Jiao}
(yjiao001@ntu.edu.sg) is currently a Ph.D. student in the School of Computer Science and Engineering, Nanyang Technological University, Singapore. He received a B.S. degree from the College of Communications Engineering, Nanjing, China, in 2013. His research interests include resource management in the Internet of Things and economics of big data.
\end{IEEEbiographynophoto}

\begin{IEEEbiographynophoto}
{Dusit Niyato}
[M'09--SM'15--F'17] (dniyato@ntu.edu.sg) is currently an Associate Professor in the School of Computer Science and Engineering, at Nanyang Technological University, Singapore. He received B.Eng. from King Mongkuts Institute of Technology Ladkrabang (KMITL), Thailand in 1999 and Ph.D. in Electrical and Computer Engineering from the University of Manitoba, Canada in 2008. His research interests are in the area of energy harvesting for wireless communication, Internet of Things (IoT) and sensor networks.
\end{IEEEbiographynophoto}

\begin{IEEEbiographynophoto}
{Ping Wang}
[M'08--SM'15] (wangping@ntu.edu.sg) received the Ph.D. degree in electrical engineering from University of Waterloo, Canada, in 2008. Currently she is an Associate Professor in the School of Computer Science and Engineering, Nanyang Technological University, Singapore. Her current research interests include resource allocation in multimedia wireless networks, cloud computing, and smart grid. She was a corecipient of the Best Paper Award from IEEE Wireless Communications and Networking Conference~(WCNC) 2012 and IEEE International Conference on Communications~(ICC) 2007.
\end{IEEEbiographynophoto}
\vfill

\begin{IEEEbiographynophoto}
{Derek Leong} (dleong@i2r.a-star.edu.sg) received the B.S. degree in electrical and computer engineering from Carnegie Mellon University in 2005, and the M.S. and Ph.D. degrees in electrical engineering from the California Institute of Technology in 2008 and 2013, respectively. He is a scientist with the Smart Energy and Environment cluster at the Institute for Infocomm Research~(I\textsuperscript{2}R), A*STAR, Singapore. His research and development interests include distributed systems, sensor networks, smart cities, and the Internet of Things.
\end{IEEEbiographynophoto}

\begin{IEEEbiographynophoto}
{Zhu Han}
[S'01--M'04--SM'09--F'14] (zhan2@uh.edu) received the B.S. degree in electronic engineering from Tsinghua University, in 1997, and the M.S. and Ph.D. degrees in electrical engineering from the University of Maryland, College Park, in 1999 and 2003, respectively. From 2000 to 2002, he was an R\&D Engineer of JDSU, Germantown, Maryland. From 2003 to 2006, he was a Research Associate at the University of Maryland. From 2006 to 2008, he was an assistant professor in Boise State University, Idaho. Currently, he is a Professor in Electrical and Computer Engineering Department as well as Computer Science Department at the University of Houston, Texas. His research interests include wireless resource allocation and management, wireless communications and networking, game theory, wireless multimedia, security, and smart grid communication. Dr. Han received an NSF Career Award in 2010, the Fred W. Ellersick Prize of the IEEE Communication Society in 2011, the EURASIP Best Paper Award for the Journal on Advances in Signal Processing in 2015, several best paper awards in IEEE conferences, and is currently an IEEE Communications Society Distinguished Lecturer.
\end{IEEEbiographynophoto}
\vfill

\end{document}